\documentclass[12pt,letterpaper,english]{aastex63}

\usepackage{geometry}
\usepackage{setspace}
\usepackage{enumitem}

\begin{document}

\title{Probing the Plasma Tail of Interstellar Comet 2I/Borisov}

\correspondingauthor{P K Manoharan}
\email{mano@naic.edu, Periasamy.Manoharan@ucf.edu}

\author[0000-0003-4274-211X]{P K Manoharan}
\affiliation{Arecibo Observatory, University of Central Florida,
HC3 Box 53995, Arecibo, PR 00612, USA}

\author{Phil Perillat}
\affiliation{Arecibo Observatory, University of Central Florida,
HC3 Box 53995, Arecibo, PR 00612, USA}

\author{C J Salter}
\affiliation{Green Bank Observatory, P.O. Box 2, Green Bank, WV 24944, USA}
\affiliation{Arecibo Observatory, University of Central Florida,
HC3 Box 53995, Arecibo, PR 00612, USA}

\author{Tapasi Ghosh}
\affiliation{Green Bank Observatory, P.O. Box 2, Green Bank, WV 24944, USA}

\author{Shikha Raizada}
\affiliation{Arecibo Observatory, University of Central Florida, 
HC3 Box 53995, Arecibo, PR 00612, USA}

\author{Ryan S Lynch}
\affiliation{Green Bank Observatory, P.O. Box 2, Green Bank, WV 24944, USA}

\author{Amber Bonsall-Pisano}
\affiliation{Green Bank Observatory, P.O. Box 2, Green Bank, WV 24944, USA}

\author{B C Joshi}
\affiliation{National Centre for Radio Astrophysics, Tata Institute of 
Fundamental Research, Pune 411007, India}

\author{Anish Roshi}
\affiliation{Arecibo Observatory, University of Central Florida, 
HC3 Box 53995, Arecibo, PR 00612, USA}

\author{Christiano Brum}
\affiliation{Arecibo Observatory, University of Central Florida,
HC3 Box 53995, Arecibo, PR 00612, USA}

\author{Arun Venkataraman}
\affiliation{Arecibo Observatory, University of Central Florida, 
HC3 Box 53995, Arecibo, PR 00612, USA}

\begin{abstract}

We present an occultation study of compact radio sources by the plasma
tail of interstellar Comet 2I/Borisov (C/2019 Q4) both pre- and
near-perihelion using the Arecibo and Green Bank radio telescopes. The
interplanetary scintillation (IPS) technique was used to probe the
plasma tail at P-band (302--352 MHz), 820 MHz, and L-band (1120--1730 MHz). 
The presence and absence of scintillation at different perpendicular 
distances from the central axis of the plasma tail suggests a narrow tail 
of less than 6~arcmin at a distance of $\sim$10~arcmin ($\sim$$10^6$~km) 
from the comet nucleus. Data recorded during the occultation of B1019+083 
on 31 October 2019 with the Arecibo Telescope covered the width of the 
plasma tail from its outer region to the central axis.
The systematic increase in scintillation during the occultation provides
the plasma properties associated with the tail when the comet was at its
pre-perihelion phase.  The excess level of L-band scintillation indicates
a plasma density enhancement of $\sim$15--20 times that of the background
solar wind. The evolving shape of the observed scintillation power
spectra across the tail from its edge to the central axis suggests a
density spectrum flatter than Kolmogorov, and that the plasma-density
irregularity scales present in the tail range between 10 and 700 km.
The discovery of a high-frequency spectral excess, corresponding to
irregularity scales much smaller than the Fresnel scale, suggests the
presence of small-scale density structures in the plasma tail, likely
caused by interaction between the solar wind and the plasma environment
formed by the comet.

\end{abstract}


\section{Introduction}

Cometary nuclei predominantly consist of highly concentrated primordial
ices, rock and dust that originated in the very outer regions of a
planetary system. They can thus supply information on the early history
of such systems. As they approach the Sun, the nuclei of solar-system
comets develop gravitationally-unbound atmospheres, or comae, surrounding
their nuclei.  These comae have diameters many times those of the nuclei,
which are typically $\sim$1~--~10~km. Moreover, spectacular tails, formed
as comets approach the Sun, are much longer and wider than the comae,
sometimes extending up to 1 AU in
length. Investigations of these tails are important for understanding
the interplanetary processes involved as a comet passes close to the Sun.

Two types of comet tails are formed. Firstly, a dust (or
type-2) tail caused by solar radiation pressure, combined with
magneto-hydrodynamic forces, which triggers sublimation processes
beneath the outer surface of the coma, followed by the supersonic flow
of dust particles from the cometary coma/nucleus.  A second plasma
(or ion) tail, also known as the type-1 tail, is caused by the complex
interaction between the solar wind and the coma through a combination
of photo-dissociation of molecules via solar extreme-ultraviolet
radiation and charge exchange with the energetic solar-wind plasma. The
charged particles gyrate around the frozen-in interplanetary magnetic
field and move along the field lines in the anti-solar direction.
The plasma tails have opening angles of a few degrees, and are mostly
observed to be longer than the dust tails. In fact, the actual size
of a plasma tail can be much longer than the structure that is visible
optically (\citealt{brandt1968}).  A plasma tail contains density
irregularities that flow out from the coma, and the associated electron-density 
inhomogeneities scatter the plane radio wave-front from a
compact background radio source passing through it causing scintillation
to be observed in the intensity of the radio source.
(e.g., \citealt{brandt1968}; \citealt{biermannetal1967}; \citealt{royetal2007}).

The measured scintillations as the line of sight to a radio source crosses the 
plasma tail of a comet can be used to study the properties of that tail. 
Such studies successfully probed the enhancement of scintillation caused
by the tail plasma of several solar-system comets at different distances from 
the Sun. In particular, {\it Comet Wilson 1987} and {\it Schwassmann-Wachmann 3-B} 
showed the presence of large-scale ($\sim$500 km) and small-scale ($\lesssim$50 km) 
density structures at the edge of the tail and near the tail axis respectively 
(\citealt{1975Ap&SS..37..275A}; \citealt{1986Natur.322..439A};
\citealt{1987MNRAS.229..485H}; \citealt{slee1990}; \citealt{royetal2007}).
Similarly, ``Interplanetary Scintillations'' (IPS) occur when the
continuum radio emission from a distant compact radio source passes
through the small-scale electron density inhomogeneities in the ionized
solar-wind plasma. Although originally much used for detecting compact
components in distant radio sources, IPS has also long been employed to
study both the solar wind, and ``coronal mass ejections'' moving rapidly
outward from the Sun 
(e.g., \citealt{kojima1987}; \citealt{coles1995}; \citealt{manoetal2001}).

Comet 2I/Borisov was only the second interstellar object known to have
entered the solar system.  The first such object, `Oumuamua' (1I/2017~U1), 
was a tiny, $\leq$1~km~$\times$ ~$\sim$100~m, elongated, asteroid-like 
object that appeared inactive even at a perihelion distance of 
$\sim$35$\times 10^6$~km~(0.23~AU). The highly-eccentric, hyperbolic orbits 
of these two objects, and their high inclinations to the ecliptic plane,
demonstrates that they have an extra-solar system origin. In contrast
to 1I/2017 U1, although also small, 2I/Borisov was an active object. Well 
before perihelion at $\sim$2~AU it manifested an elongated coma, indicating
the occurrence of jets, with the presence of cyanide and other molecules 
typically detected from solar-system comets (e.g., \citealt{manzinietal2020}; 
\citealt{cordiner2020}). As 2I/Borisov approached the Sun, solar 
radiation heated its coma, resulting in the formation of a tail, which grew 
steadily in length. In late October 2019, the observable dust tail stretched 
out from the 1~km-sized nucleus to a radial size of $\sim$2$\times 
10^5$~km, corresponding to an angular distance of more than 2 arcmin.
Figure 1 displays an image of 2I/Borisov
taken with the W.M. Keck Observatory Low-Resolution 
Imaging Spectrometer on 24 November 2019, when the comet was 2 AU from the 
Sun, along with an image of the Earth to show the size scale. 

Polarimetric observations of 2I/Borisov suggested a remarkably smooth, 
pristine coma with a high concentration of carbon monoxide, that had likely 
never interacted with the solar wind of either the Sun or any other star 
(e.g., \citealt{bagnulo2021}). Such comets are particularly interesting 
because their material is presumably the same as when our solar system 
and the extra-solar system of 2I/Borisov were formed. Understanding their 
properties, including the formation of their tails, is important.
At the Arecibo and Green Bank Observatories, we took advantage of the
passage of the plasma tail of 2I/Borisov in front of a number of compact
extragalactic radio sources to attempt investigation of the properties 
of its plasma tail using radio scintillation observations, at P-band 
(302--352 MHz), 820 MHz, and L-band (1120--1730 MHz). These 
observations were unique in determining the dimensions of the plasma tail of
2I/Borisov when it was $\sim$2 AU away from the Earth, assessing the 
level of its density fluctuations, and deriving the spectrum of its 
density turbulence, as well as the associated density scale sizes 
existing across the tail from the edge to the central axis. The wide 
bandwidths employed in these observations led to high sensitivity, 
which was an essential requirement in probing the weak level of 
density fluctuations associated with the small-scale density structures, 
$\lesssim$10 km, near the axis of the plasma tail.  In Section 2 we 
briefly summarize the theory of scintillations introduced into the radio 
emission from a compact radio source by an intervening plasma
screen, and discuss the parameters of the screen that can be derived
from single-dish observations.  Section 3 details the orbit of Comet
2I/Borisov, and source selection for our observations.  Section 4 provides 
detailed information of the observations and data analysis. It also
presents the results, which are discussed in terms of what they tell us 
about the plasma tail of 2I/Borisov. Section 5 discusses these results and 
compares them with our existing knowledge of solar-system comet plasma tails.

\section{Remote Sensing by Scintillation Measurements}

\subsection{Scintillation Index}

Electron density irregularities, ($\delta n_e$), 
in the out-flowing solar wind plasma, causes scattering of radio
waves, which together
with the motion of the irregularities across the line of sight to a
compact radio source, cause fluctuations in the apparent source intensity. 
These so-called ``interplanetary scintillations'' (IPS) are quantified by 
the fluctuations of the intensity about its mean,
${\rm \delta I(t) = I(t) - \left < \: I(t) \:\right >}$,
where {\rm I(t)} and ${\rm \left < \: I(t) \:\right> }$ are the source intensity
at time, {\rm t}, and its average level.
The scintillation index, {\it m}, is the rms of the intensity fluctuations 
normalized by the mean source intensity 
(e.g., \citealt{hewish1964}; \citealt{mano1993}).  
Scattering is considered to be ``strong'' if the rms
fluctuations imposed on a passing radio  wavefront by the medium is
$\Delta \phi > 1$\, rad. Otherwise, the scattering is considered to be
``weak''. 
For the line-of-sight to an ideal point source, the weak-scattering
regime sets in at meter wavelengths at a perpendicular heliocentric
distance to the line of sight to a radio source of $\sim$0.2~AU, beyond
which the value of {\it m} decreases  (e.g., \citealt{cohen1969};
\citealt{mano1993}). This ``strong-to-weak scattering transition'' 
region moves closer to  the Sun at higher observing frequency. 
In the weak-scattering regime, the overall effect on the emission 
from a source passing 
through the solar wind 
can be calculated by linearly adding the contributions from several thin layers 
(the first Born approximation). Additionally, in such a situation, 
the scintillation index $m$ is proportional to the plasma density fluctuations 
($\delta n_e$), which in turn is related to the density ($n_e$) of the solar wind (e.g., 
\citealt{Colesharmonetal1978}; \citealt{celnikier1987};  
\citealt{mano1993}; \citealt{asai1998}).

\subsection{The Temporal Power Spectrum}

From the Fourier transform of the time-series of the intensity fluctuations 
one can derive the temporal power spectrum of the scintillations, $P(f)$. 
If $C^{2}_{\delta n_e}(R)$ is the scattering level of the solar wind and
$\Phi_{\delta n_e}(\kappa,z)$ the spatial spectrum of density turbulence,
where $\kappa$ is the spatial wavenumber, and the line of sight to the
source is along the $z$ direction, then $P(f)$ can be given by (e.g., 
\citealt{armstrong1978}; \citealt{mano1990}),
\begin{eqnarray}
P(f) & = & (2\pi r_e \lambda)^2 \int_{\rm observer}^{\infty} \frac{{\rm d}z}{|V_p(z)|}
       \int^{+\infty}_{-\infty} {\rm d}\kappa \;\; 
       C^{2}_{\delta n_e}(R) \times \Phi_{n_e}(\kappa,z)  \times 
      F_{\rm filter}(\kappa, z)  \times  F_{\rm source}(\kappa, z)\, .
\end{eqnarray}
In this equation, $\lambda$ is the observing wavelength and $r_e$
is the classical electron radius. 
The transverse velocity $V_p(z)$ is in the  orthogonal $x$-$y$ plane. 
Since the drift rate of density irregularities across the line of
sight to a radio source causes intensity fluctuations on the ground,
the transverse speed, $V_p (z)$ and the spatial wavenumber, $\kappa$,
are related to the temporal frequency by $2\pi f\, = \, \kappa \times
V_p (z)$.  Thus the width of the temporal power spectrum is directly
proportional to the projected speed of the density irregularities
perpendicular to the line of sight \citep{mano1990}.

The Fresnel propagation filter, 
$F_{\rm filter}(\kappa, z) = 4\sin^2 (\kappa^2 z\lambda/4\pi)$, attenuates for
wavenumbers $\kappa_f \, \lesssim \, \sqrt{(2\pi/\lambda z)}$, but does
not alter the shape of the temporal power spectrum at large wavenumbers
(i.e., at small spatial scale lengths, $1/\kappa~<~1/\kappa_f$)
(\citealt{manoetal1994JGR}; \citealt{mano2000}). 
In Equation~(1), the integration includes density irregularities of all
scale sizes present in the turbulent medium.  As the square root of the integral of
the temporal power spectrum ($\sqrt{\int P(f)\, {\rm d}f}$) equals the rms of
the intensity fluctuations, it can also be normalized by the mean intensity
of the source, ${\rm <I>}$, to yield the scintillation index, $m$  
\citep{mano1993}.  At a given 
heliocentric distance, a compact source scintillates more than an 
extended one and its brightness 
distribution,  $F_{\rm source}(\kappa, z)$, not only causes an overall
reduction in scintillation but also attenuates the high-frequency end 
of the power spectrum representing scales much smaller than the Fresnel 
radius  (\citealt{coles1978}; \citealt{mano1990}).

In the case of IPS caused by the solar
wind, the scattering power falls steeply with distance from the Sun,
$C^{2}_{\delta n_e}(R)~\propto~R^{-4}$ \citep{mano1993}, while the spatial
spectrum of density irregularities has a power-law form, $\Phi_{\delta
n_e} (\kappa) \, = \, \kappa ^{\alpha}$,  with a dissipative scale size
(i.e., inner scale or cut-off scale size).  Due to the directional
flow of plasma, the electron density spectrum can be anisotropic,
reflecting the anisotropy in the associated turbulence, leading to
considerable deviation from the scintillation properties of an isotropic
distribution, $\Phi_{\delta n_e}(\kappa) = \left (\kappa_{x}^{2}
+ \frac{\kappa_{y}^{2}}{AR^2} \right )^{\alpha /2}$, where AR is the
axial ratio of the density irregularities (e.g., \citealt{armstrong1990};
\citealt{yama1998JGR}).  The three-dimensional power-law index, $\alpha$,
varies around the ``Kolmogorov'' value of $\alpha = -11/3$, within the 
range of --2.5 to --4, depending upon the heliocentric distance of 
the solar wind that is probed and its source region on the Sun  
(e.g., \citealt{ColesHarmon1989}; \citealt{manoetal1994JGR}; 
\citealt{yakovklev2018}).

\subsection{Scintillation by a Comet Plasma Tail}

For scintillations produced by a comet plasma tail, the thickness of the 
turbulent screen is determined by the effective path length through the 
tail where the line of sight to the source crosses it.  Naturally, the 
line of sight also includes scattering associated with the ambient solar
wind along its path. %(i.e., above as well as below the tail).  
For weak scattering, the variances of comet-tail and
solar-wind scintillations simply add, i.e., their power spectra sum.
As the solar-wind scattering falls off rapidly with distance from the Sun
($C^{2}_{\delta n_e}(R)~\propto~R^{-4}$), for a large solar elongation of
the source, the effect of the solar wind on the observed scintillation
will be less significant and the contribution from the comet ion tail
dominates the measured scintillation. For a comet tail, the rate
of intensity fluctuation is determined by the concentration of plasma
in it and the transverse velocity of the irregularities. The shape of
the temporal power spectrum is the combined effect of the flow speed of
the plasma away from the nucleus, the irregularity scale sizes, and the
level of scattering power. In the present study, the comet was located
at distances of $\gtrsim$2~AU from the Earth, for which the associated size
of the Fresnel radius is about 40\% larger than that expected for the
solar-wind scintillation. Since the low-frequency part of the temporal
spectrum is determined by the effect of large-scale plasma irregularities
close to the Fresnel radius, in the present case, the plasma associated 
with the comet tail is expected to move the Fresnel knee toward the low 
temporal frequencies, and to increase the scintillation at the low-frequency
part of the spectrum.

\section{The Orbit of Comet 2I/Borisov and the Selection of Occultation
Sources}

The path of Comet 2I/Borisov from mid-October to mid-December 2019 is
shown in Figure 2. For each day, the orbit of the comet was obtained
with a time resolution of 30~min from the ``Solar System Dynamics Group,
Horizons On-Line Ephemeris System", provided by the Jet Propulsion
Laboratory \url{(http://ssd.jpl.nasa.gov/)}.  During this period, the
Sun-comet and Earth-comet distances lay between 2.5 -- 2.0~AU and 2.9
-- 2.4~AU respectively.

In Figure 2, the positions of the radio sources selected for observation
as the plasma tail of the comet passed over them are also marked along
the path of the comet. These sources were selected from the Texas and
Molonglo catalogs at 365 and 408 MHz (\citealt{large1981}; 
\citealt{texas1996}), respectively, plus the results 
of IPS measurements made with the Ooty Radio Telescope at 327 MHz (i.e., 
\citealt{manolfru2009}; \citealt{mano2012}).  
Since only limited information was available on the
length and width of the comet plasma tail, we selected sources in the sky
plane whose lines of sight passed within $\sim$12~$\times$~12~arcmin
of the coma of the comet, namely B0957+142, B1019+083, B1023+067, and
B1129--202, located on the tail side of the comet on 20 and 31 October, 03
November, and 11 December 2019 respectively. The observations of the first
3 sources were made with the Arecibo 305-m Telescope, while B1129--202
was observed with the Green Bank 100-m Telescope. 

Previous extensive monitoring of IPS for sources B0957+142 (4C14.35) and 
B1023+067 (3C243) was available from the Ooty Radio Telescope, operating 
at 327 MHz 
(\citealt{swarup1971}; \citealt{mano1995SoPh}; \citealt{manolfru2009}).
For example, a large number of IPS observations of B0957+142, nearly 
300 observations between 1992 and 2016 at different elongations, allowed 
estimates of a scintillation component size, $\Theta$~$\approx$~150~mas, 
and flux density, $\Delta$S~$\approx$~700 mJy. B1023+067 was monitored 
about 100 times between 1989 and 2016, these observations providing
$\Theta$~$\approx$~100~mas and $\Delta$S~$\approx$~600~mJy. 

For the solar wind monitoring at Ooty, 
strong sources (S$_{\rm 327}$~$\gtrsim$~2~Jy) were mostly employed and compact
component details were not available for the weak sources B1019+083 and 
B1129--202. However, after the present comet observations, we requested IPS 
observations at Ooty and observed B1019+083 during July 2020 and October 2021 
(Section 4.1, Figure 3). We did not get an opportunity to similarly observe B1129--202. 
The parameters of these sources are listed in Table 1, which also summarizes 
some of the relevant observational parameters.

\section{Observations, Data Analysis and Results}

\subsection{Ooty Control IPS Observations of B1019+083 at 327~MHz}

To provide detailed information on B1019+083 at P-band (302--352 MHz) (also illuminating 
the situation at 1400 MHz), 327-MHz IPS observations were requested with the 
Ooty Radio Telescope (ORT; \citealt{swarup1971}), when the source approached 
and receded from the Sun. These were made during 14--20 July 2020 and 06--07 
October 2021 respectively. The solar elongations of these observations ranged 
between 36$^\circ$ and 43$^\circ$ (i.e., corresponding to a solar radial offset 
range of 0.57--0.67 AU). On each day, on-source and off-source data were 
collected for about 15 and 5 min respectively, along with the observation of 
a flux-density calibrator. The observational method and data analysis procedures  
followed are given in \citet{mano1993} and \citet{mano2000}. The Ooty temporal 
power spectrum at $\lambda$ = 0.92 m, in the frequency range of $f$ $\approx$ 
0.1 -- 25 Hz, includes spatial wavenumbers in the range 
$0.002 < \kappa =\frac{2\pi f}{V} < 0.2$ km$^{-1}$, corresponding to solar-wind 
density turbulence scales between 5 and 500 km for a typical solar wind 
velocity of 350 kms$^{-1}$, and 
the angular size of the source, $\Theta$~(FWHM)~$>$400 milliarcsec (mas), heavily 
attenuates the scintillation. During these observations, 2 days  were affected 
by interference. Power spectra of the other 7 days are shown in Figure 3, each in 
a different color. The solar elongation of each day's observation is also 
indicated. At such large elongations, the line of sight to the radio source 
lies close to the ecliptic plane and the solar wind properties would 
be those of low heliographic latitudes.
The figure also contains an IPS model spectrum 
obtained from Equation~1. This model was computed for a solar wind speed, 
V = 350 km/s, a source size, $\Theta$~=~50 mas, and a power-law index, 
$\alpha$~=~--3.3. Moreover, the average model fit to the
observed power spectra provides
an estimate for the source size of $\Theta$~=~50$\pm$25 mas, and solar wind speeds in the 
range of 300 to 400 kms$^{-1}$, which is consistent with the near-Earth {\it in-situ} 
measurements of the solar wind speed in the periods 12--25 July 2020 and 4--9 October 
2021 (\url{https://cdaweb.gsfc.nasa.gov)}. Additionally, the observed average 
scintillation index, $m$~$\approx$~0.3, corresponds to a flux density of the 
compact component, $\Delta S$~$\approx$~500 mJy at 327 MHz.

\subsection{Observations, Data Analysis and Results at Arecibo}

IPS measurements were made 50 years ago with the Arecibo Telescope,
chiefly to identify sources with sub-arcsecond structure (e.g.,
\citealt{cohen1967}; \citealt{cohen1969}; \citealt{harris1973}).
Recently, exploiting the post-1990s expanded frequency coverage of the
telescope, (300 MHz~--~10 GHz), IPS observations were recommenced
at Arecibo to study the characteristics of turbulence in the solar-wind
plasma at a range of distances from the Sun. A full suite of analysis
and display programs have been developed for these new measurements
(\citealt{manoaas2021}).
%(\citealt{ngat2021}).  

The present study of Comet Borisov used two of the 305-m Telescope receivers,
the so-called ``P-Band'' and ``L-Band Wide'' systems, covering the frequency
ranges of 302-352, and 1120-1730 MHz respectively. The gains are $\sim$10~K/Jy, 
with both receivers, recording both orthogonal polarizations. 
The FPGA-based Mock spectrometer system was used in single-pixel mode
(see \url{http://www.naic.edu/~astro/mock.shtml}) to record the data.
At P-Band, a single Mock-spectrometer box was employed. This recorded 
53.3-MHz bandwidth of dual-polarization data centered at 327 MHz, with
each polarization divided into 1024 channels, and having an integration
time of 2~msec. At L-band, all 7 available Mock boxes were employed,
each processing an 80-MHz bandwidth of dual-polarization data. The bands
were centered at 1160, 1240, 1320, 1400, 1480, 1620 and 1700~MHz. Here,
each band was divided into 2048 channels per polarization, with the data
being sampled at an integration time of 1~msec.

The occulted sources were tracked for $\sim$2~--~2.5~hours. Off-source
data were taken before and after the on-source scans by moving the 
telescope pointing east of the source by $\sim$3~min in right ascension
and were useful to compute the source deflection, which also included 
the system gain variation. Bandpass correction was applied to every 
10~sec of data (see \url{https://www.naic.edu/~phil/masdoc.html}).
Each 80-MHz
band at L-band was split into 8~$\times$~10-MHz subbands, with each
subband having 256 channels. Individual bad channels within each subband
were identified by their high transient rms values ($>$3$\sigma$) and
flagged. For each subband, the frequency-averaged spectral density over
the remaining good channels was then calculated for each 1-msec sample,
resulting in total-power time series for 56~$\times$~10-MHz subbands of
L-band data.  These were computed for on- and off-source scans.  For the
P-Band observations, total-power time series for 5~$\times$~10-MHz
subbands were similarly computed.

Running means of 10-sec duration were subtracted from the data streams
to remove any slow variations at frequencies below 0.1~Hz, Fourier
transforms performed and temporal power spectra computed for 1-min
blocks of data for each 10-MHz subband. 
(The above running mean subtraction is also useful to remove any 
low-frequency system variations, if present, and/or the response of 
the slow drifting of a `screen' (e.g., ionosphere) located close to 
the observer. However, it will not alter the faster intensity 
fluctuations generated by the plasma tail.) 
Average 1-min power spectra were obtained for P- and L-band by averaging 
all subbands and used for the scintillation index calculation.
For display purposes, 2-min spectra were computed for L-band to 
improve the stability of the spectrum.
At high temporal
frequency, $>$10 Hz, the spectrum dropped to the level of the receiver
system noise, a constant level independent of frequency. 
This was subtracted from the on-source power spectrum to obtain the 
spectrum of the pure intensity fluctuations.  
However, in the noise-level subtracted spectrum, the fractional error
increased monotonically as the scintillation power approached the system 
noise (Figure~7(a), spectrum shown in `black'). 
In order to achieve a nearly constant fractional error over the 
usable region of the spectrum, the frequency resolution at $<$0.5 Hz was 
retained at $\sim$0.1 Hz, while at higher frequencies adjacent points 
were averaged, with the number of points averaged increasing such as to 
keep the spectral points equispaced on the logarithmic scale of the 
frequency axis (Figure~7(a), spectrum shown in `red').

\subsubsection{The Central Occultation of B1019+083 on 31 October 2019}

Figure 4 shows the path of the comet on 31 October 2019 with respect to
the target source B1019+083, projected on to the plane of the sky. The
arrow indicates the direction to the Sun, while the thin-dotted line
connects the comet nucleus and the Sun at 12 UT. On 31 October, the 
solar elongation of the source was 64$^\circ$, corresponding to a solar 
offset of 0.9 AU.

At P-band (i.e., 302-352 MHz), the source was observed for about 
one hour from $\sim$11:10 UT. During this time, the perpendicular angular 
separation between the center of the comet tail and the line of sight 
to the source decreased from about 2.5 to 1.5~arcmin.  The P-band 
temporal power spectra did not show any significant change with time 
as the line of sight to the source approached the plasma tail of the 
comet. The scintillation index values computed from the power spectra of 
1-min data stretches (i.e., ${m\,\,=\,\,\sqrt{\int P(f)\, {\rm d}f}/{\rm <I>}}$) 
are plotted in Figure 5. The error bar shown on each point is the `peak-to-peak'
variation of 5 scintillation indices obtained from the 5 individual subbands,
each of 10-MHz width.
The rather steady, low level of signal fluctuation seen throughout the 
entire hour spent at P-band suggests  that at a distance of 
$\sim$3-4$\times 10^5$ km from the nucleus of the comet, the detectable
width of the tail was $\lesssim$3~arcmin.

Between $\sim$12:25 and 13:20 UT, B1019+083 was observed with the L-band
system.  During this time, the angular separation between the axis of the 
comet tail and the line of sight to the radio source decreased
from $\sim$1~arcmin ($\sim$$10^5$ km), with the central part of the tail
expected to occult the line of sight at around 13~UT.  For each 1-min
spectrum, the scintillation index was computed as detailed above. Figure
6 shows the scintillation index as a function of time, reaching a maximum
at $\sim$13 UT.  After this time, the apparent scintillation decreased.
However, the source had reached the tracking limit of the telescope, and 
the telescope pointing moved quickly away from the source (i.e., within less 
than 15~s of time, corresponding to the size of the beam at L-band), with 
the fluctuation level rapidly dropping to a steady low value. By 13:04
UT, the source had moved right out of the telescope beam, and on
to a source-free region of the sky, providing a representative
baseline of system noise.

\subsubsection{The Evolution of the L-Band Power Spectra of B1019+083
During its Occultation}

When the L-band temporal power spectra of B1019+083 were examined
carefully, a systematic evolution could be seen in the spectral shape 
and integrated power as the line of sight to the source approached 
the central part of the plasma tail (see Figure 6). 
%as shown in Figure 7.
% In this figure, in order to get an improved 
% sensitivity and stability at the high-frequency part of the spectrum, 
% we display spectra averaged over 2-min of measurements.
Figure~7(a) displays the spectrum observed at $\sim$12:53 UT, approaching 
the peak seen in Figure 6, after subtracting the
off-level of the spectrum corresponding to the level of system
white noise. In the spectrum shown in `black', adjacent pairs of points
have been averaged, whereas in the `red' plot averaging is such that
the points are nearly equispaced on the logarithmic scale. Figure~7(a)
also includes two model spectra computed for a spherically diverging
solar-wind velocity of 360 km/s at a solar elongation of 64$^\circ$
using Equation~(1).  The `green' spectrum is for a power-law exponent,
$\alpha$~=~--2.7, while the `blue' spectrum corresponds to $\alpha$~=~--3.3.
In contrast, if two different velocity fittings are considered separately 
for the low- and high-frequency parts of the resultant spectrum, then its 
high-frequency part would scale to an abnormal high velocity, $>$2000 km/s. 
The displayed model spectra adopt an angular source size of $\Theta$~=~50~mas 
and an anisotropy parameter, AR~=~1.5. The derived spectrum from the
observations shows the narrowing 
at the low-frequency part, caused by the large Fresnel scales at a
distance of $\sim$2~AU for the plasma screen. 

Moreover, comparing the spectra from the beginning of the L-band 
observations with those near the peak of the scintillation (see Figure~7(b)~) 
shows remarkable changes. The high-frequency part of the spectrum becomes 
prominent as the central axis of the tail is approached, indicating the 
presence of small-scale density structures as well as an increase in the 
scattering power near the central axis of the plasma tail.
Figure 7(b) displays 2-min average spectra corresponding to, (i) the start 
of the L-band observations at $\sim$12:26~UT, (ii) close to the peak of the
scintillations at $\sim$12:57~UT, and (iii) at $\sim$13:17 UT, when the
telescope pointing lay outside of the plasma tail and at an off-source region. 
In this figure, UT times are marked on the spectra.  Both recent and earlier 
IPS observations at Arecibo have demonstrated that the L-band scintillation
index of selected compact radio sources (e.g., 3C237 and 3C138) drop
by a factor of $\sim$30 between a peak value at $\sim$0.07~AU and a
solar offset of 0.9~AU (e.g. \citealt{cohen1969}; \citealt{manoaas2021}),
corresponding to a reduction in spectral power of $\sim$15~dB; the solar 
elongation of the present observations correspond to a solar offset for 
the line of sight of $\sim$2~AU.
When the line of sight to the radio source lay close to the center-line
of the tail, the power spectrum -- plotted in ``red''  in Figure 7(b) --
became broad, extending up to $\sim$10~Hz, and its amplitude increased.  
This suggests increased scattering when the central
part of the tail was approached.  At frequencies above 10~Hz, the 
spectrum merges with the white-noise. Below 4~Hz, the spectrum indicates 
a power-law of slope $\alpha \approx$~--3.  A further
prominent feature seen in the temporal power spectrum is that excess
scattering power exists in the 4--9~Hz region.  Similar results have been
obtained for the solar-system comets, Schwassmann-Wachmann~3-B and Austin
(\citealt{royetal2007}; \citealt{chris2019}).  This excess power seems
to decrease with offset of the line of sight from the center of the
tail. The power spectrum derived at $\sim$12:26 UT, shown in ``blue''
in Figure 7(b), corresponding to the start of the L-band observation when
the line of sight was at the edge of the plasma tail, shows a similar,
but relatively smaller, excess power. When the beam drifted away from
the target radio source, the spectrum, shown in ``green'' in Figure
7(b), contained insignificant power 
and was essentially flat.

In order to have comparison data,  B1019+083 was observed again for
a short time at L-band on 05 November 2019. The equivalent temporal
power spectrum from this control observation is plotted in ``black''
in Figure 7(b). This spectrum is nearly flat over the full range of
temporal frequencies, at a solar elongation of 69$^\circ$.

\subsubsection{The Arecibo Observations of Two ``Near-Occulatations''}

Two other radio sources, B0957+142 and B1023+067, (both established
IPS sources at 327 MHz, see Table~1), were also observed at
Arecibo  on 20 October and 03 November 2019 respectively.
As for B1019+083, both were observed consecutively at P-band (302--352 MHz) and 
L-band (1120--1730 MHz) for about an hour
at each frequency  during their individual Arecibo meridian
transits. Both sources are fairly strong, the sizes of their 
compact components being $\sim$100--150~mas, (see Table 1). The data analysis was 
performed as described above, and the 1-min temporal power spectra were 
computed for both days. Their derived spectra were nearly flat with no 
excess signal fluctuations being apparent.

The sources were located at angular distances of $\sim$10~arcmin from
the comet coma in the tail direction. Moreover, since the actual tail occultation
times on the above days fell slightly outside of the tracking limit
of the telescope, these sources were observed when the lines of sight
were separated from the central part of the tail by 
an orthogonal distance of $\sim$6~arcmin. While the
non-detection of scintillations could indicate that the tail length was $<$10
arcmin from the comet coma, corresponding to a radial distance of
$\sim$8$\times 10^5$ km, the plasma tail was possibly
longer than 10 arcmin, but sufficiently narrow that the occultations only 
began after the sources had set at Arecibo.

\subsection{A Central Tail Occultation Observed with the GBT}

On 11 December 2019, just after the perihelion passage of Comet 2I/Borisov, 
we observed its ion tail occultation of the radio source B1129--202 using 
the 100-m Green Bank Telescope (GBT) and its PF1\_800 receiver at 820 MHz. 
The solar elongation of the source was 77$^\circ$. 
The source was located at an angular separation of $\sim$12
arcmin ($\sim$$10^6$~km) from the Comet Borisov nucleus in the anti-solar
direction.  Figure~8 displays the path of the comet on that day, as
projected on the plane of the sky, with the position of B1129--202
shown on the plot. The arrow indicates the direction to the Sun, while
the thin-dotted line connects the comet nucleus and the Sun at 11 UT.
It is to be noted that this radio source is relatively weak, its 
interpolated flux at 820 MHz being 0.37 Jy, with a spectral index of 
--0.83 between 76 and 4850~MHz using flux densities from VizieR, brought 
on to the flux-density scale of \citet{baars1977}.
% flux density at 820~MHz being 0.41~Jy, with a spectral index of 
% $\sim -0.83$ between 76 and 4850~MHz {\bf (\citealt{NED2016}) }.  
There is no $a$~$priori$ information available on the existence of any 
compact component that this source may contain.  
(Since the number of radio sources occulted by the plasma tail was 
limited, we included all possible occultation sources to the observation
list, irrespective of their angular sizes, with an additional aim of 
detecting the rare event of tail detachment, if such were to occur.) 
Starting from 09:22~UT, 
the source was observed for about 4 hr. The central part of the tail 
was expected to cross the line of sight to the source at $\sim$11~UT.

\subsubsection{Observations, Data Analysis and Results at Green Bank}

While the telescope tracked the radio source B1129--202,
two backends simultaneously recorded data in two different
acquisition modes. These were: (i) the GUPPI pulsar backend (see
\url{https://safe.nrao.edu/wiki/bin/view/CICADA/GUPPISupportGuide}),
recording total-intensity (I), dynamic spectra in pulsar-search mode
with 2048 channels over a bandwidth of 200~MHz every 1-msec, and
(ii) the VEGAS spectrometer (see \url{https://www.gb.nrao.edu/vegas/}),
recording spectra of the orthogonal linear polarizations (XX \& YY)
with 32768 channels over a 187.5-MHz bandwidth every 100~msec.
Each ``scan'' was of 58~min duration. Before each scan, the total-power
level of the entire signal chain was re-adjusted. Four such scans were
made beginning at 09:22~UT, with the fourth scan lasting only about
30~min. However, due to initial set-up issues, VEGAS recording began
only from the second scan at 10:26~UT.

Early in the data-reduction process, it became clear from the temporal
power spectra of the VEGAS data that both the XX and YY polarizations
were affected by an ever-present, quasi-periodic signal generated
by the cryogenic vacuum pump of the receiver system. This had a
well-defined 0.8333~sec periodicity, producing a strong spike in the
temporal power spectra at a frequency of 1.2~Hz, plus a number of higher
harmonics. This unwanted signal was much stronger in the YY polarization
than in the XX channel.  In addition, the VEGAS time series of the
YY polarization signal exhibited an extended episode lasting in total
$\sim$30~min close to the expected time of occultation by the comet ion
tail. During this time interval, the signal level steadily decreased by
$\sim$10\%, subsequently recovering on a similar time scale to its former
level. No such deviation was seen in the VEGAS XX-polarization signal,
and it is hence believed that this ``glitch'' in the YY polarization was
purely instrumental. Both this extended ``downward excursion'', and the
quasi-periodic interference from the cryogenic pump, were also seen in the
Stokes-I time series of the GUPPI data. Hence for the GBT data, only the
signal from the VEGAS XX polarization was utilized for further analysis.

Data analysis at Green Bank proceeded as follows:
\begin{enumerate}[label=\arabic{enumi}]

\item For all three scans recorded by VEGAS, average XX-polarization
spectra were visually inspected, and channels that were heavily
RFI-affected were noted. From this, six independent 10-MHz wide frequency
ranges were selected that were essentially RFI-free.

\item A fifth-order polynomial fit was made to each 58-min (and $\sim$30~min 
for the final scan) scan-averaged spectrum, using just the data within the 6
$\times$ 10-MHz RFI-free bands to construct an average bandpass spectrum. 

\item For each 0.1-sec VEGAS XX-polarization data sample, the recorded
spectrum was normalized by the average bandpass spectrum (obtained from 
step \#2), and the 6 $\times$ 10-MHz spectral bands were each averaged in 
frequency to create 6 $\times$ total-power, 10-MHz bandwidth, continuum 
time series.

\item From these time series, for every 2~min, temporal power spectra
were computed, and averaged over the six independent frequency sets. This
produced final temporal power spectra, (effectively averaged over 60 MHz).
This avoided band-width smearing of possible scintillation effects
had we just produced a simple 60-MHz band-width averaged time series
before calculating the corresponding power spectra.

\item Finally, measuring the area under these power spectra between
temporal frequencies of 0.1 and 0.9~Hz, (thus avoiding the effects of the
cryogenic pump, very low-frequency contamination caused by any low-level
ionospheric scintillation and/or slow drifts), provided a measure of the
power contained in fluctuations that could be caused due to non-instrumental
effects.

\end{enumerate}

The measured fluctuation power, ${\int P(f)\, {\rm d}f}$, 
is shown plotted against time in
Figure 9. This shows no significant change during the first 2~hr during
which the VEGAS data was acquired, the time during which the cometary ion
tail had been predicted to occult the background radio source. However,
the final, shorter, 30-min scan indicates a sudden increase in the
``fluctuating power''. In respect of this, we note that on Dec~11 2019 the
Sun rose at Green Bank at 12:26 UT, exactly 2~min after the start of that
scan. We conclude that solar emission in the far sidelobes of the antenna
could possibly have caused the sudden increase in fluctuations registered
in the time-series, rather than this being due to  scintillations of
the source emission caused by the ion tail of the comet.

For a single polarization, with a 60-MHz bandwidth, total-power channel, 
sampled at 0.1 sec, the GBT 800-MHz receiver (with a System Equivalent 
Flux Density (SEFD) of 10 Jy) would theoretically yield a rms noise
of ($\rm{SEFD}/\sqrt{\beta \tau}$)~$\sim$~4.1~mJy. Any fluctuations in
excess of about 5 $\times$ this noise would be expected to be caused via a
physical process such as IPS, (or maybe ionospheric scintillation for time
scales on the order of 10's of sec). At a given solar elongation, the 
scintillation of a source is detectable when its compact component flux 
density is at least $\sim$20 mJy (\citealt{mano1995SoPh}). 

% Depending upon the solar elongation,
% compact sources show scintillations from 100\% to $\lesssim$10\% (e.g.,
% \citealt{mano1995SoPh}).  
% Thus, for a point source to show a Scintillation
% Index of at least 10\% for these GBT observations, it should have a
% compact component of at least $\sim$200~mJy flux density for a 5-$\sigma$
% detection. We do not have any independent verification of the existence
% of such a component in B1129-202. [MANOHARAN, WE ARE WAY OUTSIDE OUR
% LEVEL OF COMPETENCE HERE! CAN YOU ADD SOMETHING MEANINGFUL GIVEN THE
% GBT OBSERVATIONS AND THE SOLAR ELONGATION OF 77 DEG HERE PLEASE?]

The occultation observation of B1129-202 with the GBT on 
11 December 2019 and the control observation on 04 January 2020, 
respectively at solar elongations 77$^\circ$ and 99$^\circ$, showed 
the integrated scintillation spectral power, at frequencies below 1 Hz, 
equivalent to a flux density of about 20 mJy. 
From the 2.5-arcsec resolution image at 3 GHz of the Very Large
Array Sky Survey (VLASS; \citealt{gordon2021}), B1129-202 is seen to
consist of a pair of diametrically opposed lobes with hotspots at their
outer edges, and similar intensities. Given the lack of significant
central component, the IPS that we detect has to come from compact 
structure in these hotspots, which are separated by 1.3 arcmin, 
about one-tenth of the GBT's half-power beamwidth.  
However, in view of the lack of increased signal fluctuations at the
anticipated time of the occultation, the quasi-coincidence between the
apparent increase in the level of fluctuations and sun-rise at 
$\sim$12:26~UT, we refrain from using the GBT data from further
discussion here. 

\section{Discussion and Summary}

Scintillations of the radio emission from the source B1019+083 seen
through the plasma tail of interstellar Comet 2I/Borisov were detected at
Arecibo at 1400~MHz on 31~October 2019. At that epoch the radio source
had an angular separation from the comet nucleus of 5~arcmin, $\sim$$4 \times
10^{5}$~km, in the anti-solar direction.

Several other studies had earlier been successful in detecting
scintillations of the emission of compact radio sources seen
through the plasma tails of solar-system comets at a variety of
observing frequencies, (e.g.  \citealt{1975Ap&SS..37..275A};
\citealt{1986Natur.322..439A}; \citealt{1987MNRAS.229..485H};
\citealt{slee1990}; \citealt{royetal2007}). All have involved short
transit times for the compact radio sources through the tails, and
effectively provide point measurements of the plasma properties of the
tails. The density variations inferred from the excess scintillations
caused by the plasma tails have covered a range of rms densities of $\sim$1 --
10 cm$^{-3}$, which were above the values of the background solar-wind
density. In the present case, the average solar wind density at the
near-Earth orbit was around 5 cm$^{-3}$ for the period between 30 October
and 15 December 2019 (\url{https://cdaweb.gsfc.nasa.gov)}, and for a
typical spherically symmetrical expansion of the solar wind, this would
be lower by about a factor of 4 at the comet distance of $\sim$2 AU from
the Sun. 
In respect of a relationship between the density ($n_e$) and its 
fluctuations ($\delta n_e$) in the solar wind, the ISEE-1 and -2 space 
propagation experiments showed that $\delta n_e$~$\propto$~$n_e$, while  
high-time resolution measurements by the ISEE-3 spacecraft demonstrated
that $n_e$~$\propto$~$\delta n_e^{0.85}$ (\citealt{celnikier1987}; 
\citealt{zwickl1988}). In general, since the level of IPS at L-band 
is lower by an order of magnitude than that at 327 MHz (\citealt{manoaas2021}),
the excess levels of scintillation observed on 31 October 2019, when 
the comet was located at a distance $\sim$2 AU, compared with the Ooty 
IPS (Figure 3), suggest the density enhancement in the plasma tail to 
be in the range of $\sim$15 -- 20 times higher than that from the 
background solar wind. 

Figure 10 shows the expected level of scintillation, caused by the rms 
electron density variations of the background solar wind, along the line-of-sight 
between the observer and a distance of 3.2 AU. It includes the position of 
crossing of the plasma tail at 2.4 AU. The continuous curve represents the 
scintillation weighting function at a solar elongation of 64$^\circ$ for 
an observing frequency of 1420 MHz, and also includes the effect of an 
angular source size of 50~mas. The observed scintillation index is
the integration of the contribution of ${\delta n_e(z)}$ at each point
along the line of sight, 
$m^2 \,\sim\, \int \left[ \delta_{n_e}(z)\right]^2 \, {\rm d}z$.
The dotted line indicates the maximum contribution by the background
solar wind to the observed scintillation index, the excess above this 
being the scintillation due to the plasma tail of the comet. In Figure 6,
it is to be noted that the increase in scintillation between the background
solar wind at 0.9 AU (i.e., $\epsilon$~=~64$^\circ$) and near the central 
axis of the tail is $\sim$4 times or more. However, the scintillation due 
to the solar wind changes with the solar offset ($R$), as $m\,\sim\,R^b$ 
and the radial dependence index $b$ ranges between -1.5 and -1.6 and does 
not significantly change with the observing frequency (\citealt{cohen1969}; 
\citealt{mano1993}; \citealt{manoaas2021}). Thus a normalization of 
scintillation of the background solar wind at the distance of the tail would 
lead to an increase in the integrated $\delta_{n_e}$ in the range of 15 -- 20 
times and more. 

For the present observations, the contributions of scintillations
at 1400~MHz from IPS at the tangential heliocentric distance of
R~$\approx$~0.9~AU for radio source B1019+083 on 31~Oct 2019 are expected
to be insignificant and they can add little power to the observed temporal
spectra. For example, the observed temporal spectra on 31~October were
well above the 10~dB level (see Figure 7a), extending to frequencies $>$4~Hz,
and as the line of sight to the source approached the center-line of the tail,
additional features are seen at frequencies $\geq$5 Hz.  When we carefully
examine the shape of the spectra derived for the 1400-MHz band at temporal
frequencies $\leq$4~Hz, (see Figure~7b), the spectral characteristics
resemble those of a typical IPS model spectrum, although with a much
higher level of scintillation than is expected from IPS. It is likely
that the observed spectral shape is the result of scattering caused by
two types of prominent plasma density inhomogeneities. In 
the spectral region below 4~Hz, it is possible that the scintillations
are caused by plasma density irregularities via the interaction of
the solar wind with the comet coma. The transfer of energy from the
solar wind to the cometary coma, though this interaction is an open
question (e.g., \citealt{brandt1968}; \citealt{biermannetal1967};
\citealt{gombosi2015}), causes the comet itself to represent a source of
interplanetary plasma which becomes dominant as the comet approaches
the Sun (e.g., \citealt{gombosi2015}). In the neighbourhood (or
vicinity) of solar-wind interaction, the plasma flow characteristics
are largely controlled by the properties of the solar-wind flow, and
scintillations resembling those of IPS are scaled to an enhanced level
by strong localized density perturbations. In fact, this region of the
spectrum below 4~Hz dominates the overall scintillation index, and
also displays a well-formed Fresnel knee. It also shows the signature
of anisotropy, the `rounded knee', with axial ratio of $\approx$~1.5, 
possibly influenced by the directed plasma flow (see Figure 7a).

The temporal spectrum also shows broad, prominent spectral features in
its high-frequency part, $\gtrsim$5~Hz, in which the contribution from the  
background solar wind at a solar offset of 0.9 AU is expected to be 
insignificant for L-band measurements. These are probably caused by 
density inhomogeneities present in the central part of the plasma tail.
The images of Comet 2I/Borisov obtained with the Hubble Space Telescope
and by the W.~M.~Keck Observatory have been examined for 16 November
to 09 December, including the image shown in Figure 1. The Hubble Space
Telescope also provided images following the comet's  perihelion
passage on 08 December, after which the size of the dust tail marginally
increased.
However, as demonstrated by the images of several solar-system comets,
the plasma tail is generally well collimated compared to the dust tail,
and follows the direction of the interplanetary magnetic field (e.g.,
\citealt{gombosi2015}).  The high-frequency spectral features are likely
due to one, or a combination, of the following factors: (i) an associated
spatial density spectrum much flatter than the Kolmogorov spectrum, (ii)
a highly directional flow of plasma, resulting in an anisotropic spectrum,
(iii) a high-speed flow, or (iv) strong scattering conditions leading
to a broad spectrum.  
Since the level of scintillation is low and the spectrum is not 
broadened at its low-frequency part, a strong scattering situation is
unlikely (e.g., \citealt{mano2010a}).  
The gradual changes in the shape of the spectrum, and the high-frequency 
part becoming prominent as the central axis of the tail is approached, 
indicate the presence of small-scale density structures near the central 
axis of the plasma tail. 
A plasma flow acceleration leading to velocities much higher than those 
of the solar wind is implausible in the cometary environment, 
because an overall shift in the spectrum should then be seen. Moreover, 
if the high-frequency broadening alone is considered to be due to the 
scaling of velocity, it would relate to an unrealistic flow velocity, 
$>$2000 km/s. 
The presence of magnetic fields, which in general pervade
the turbulent interplanetary plasma, and their strong interactions,
can ultimately lead to filamentary structures, knots and/or kinks in the
plasma tail. Thus, an anisotropic, flat spatial density spectrum
can be a possible cause for the broad spectral feature.  In Figure 7(a),
we show a model spectrum with a flat power-law exponent of $\alpha=$~--2.7 and
an axial ratio of 1.5. This demonstrates approximate correlation with the
high-frequency part of the spectrum.

Thus, the overall observed power spectrum of the plasma tail of Comet
2I/Borisov likely contains different types of density spectra and
shows the presence of density scales in the range of $\sim$10 -- 700 km,
which are likely associated with the interplanetary environment produced
by the comet, plus interaction with the solar wind. Section 4.2.2 discussed 
the evolution of the spectral shape, as the line of sight to the 
source B1019+083 approached the central part of the tail (see Figure 7(b)).  
In particular,  the result on the high-frequency excess, corresponding to 
irregularity scale sizes much smaller than the Fresnel scale, is similar to 
the previous studies of the two solar-system comets, Schwassmann-Wachmann 3-B 
and Austin, albeit within 1 AU of the Sun (\citealt{royetal2007}; 
\citealt{chris2019}). These results suggest the presence of small-scale 
density structures in the concentrated central part of comet tails.

\begin{acknowledgments}

The Arecibo Observatory is operated by the University of Central Florida
under a cooperative agreement with the National Science Foundation
(AST-1822073), and in alliance with Universidad Ana G. Méndez and
Yang Enterprises, Inc. The Green Bank Observatory is a facility of the
National Science Foundation (NSF) operated under cooperative agreement
by Associated Universities, Inc. The image in Figure~1 has been
obtained from the W.M. Keck Observatory, which is operated as a scientific
partnership among the California Institute of Technology, the University
of California and NASA. We also acknowledge the JPL Solar System Dynamics
Group's Horizons On-Line Ephemeris System \url{(http://ssd.jpl.nasa.gov/)}
for the comet ephemeris data and the Coordinated Data Analysis Web (CDAWeb)
service for providing solar wind data \url{(https://cdaweb.gsfc.nasa.gov/)}.

\end{acknowledgments}

\bibliographystyle{aasjournal}
\bibliography{astro_ph_version}{}

\begin{thebibliography}{}
\expandafter\ifx\csname natexlab\endcsname\relax\def\natexlab#1{#1}\fi
\providecommand{\url}[1]{\href{#1}{#1}}
\providecommand{\dodoi}[1]{doi:~\href{http://doi.org/#1}{\nolinkurl{#1}}}
\providecommand{\doeprint}[1]{\href{http://ascl.net/#1}{\nolinkurl{http://ascl.net/#1}}}
\providecommand{\doarXiv}[1]{\href{https://arxiv.org/abs/#1}{\nolinkurl{https://arxiv.org/abs/#1}}}

\bibitem[{{Alurkar} {et~al.}(1986){Alurkar}, {Bhonsle}, \&
  {Sharma}}]{1986Natur.322..439A}
{Alurkar}, S.~K., {Bhonsle}, R.~V., \& {Sharma}, A.~K. 1986, \nat, 322, 439,
  \dodoi{10.1038/322439a0}

\bibitem[{{Ananthakrishnan} {et~al.}(1975){Ananthakrishnan}, {Bhandari}, \&
  {Rao}}]{1975Ap&SS..37..275A}
{Ananthakrishnan}, S., {Bhandari}, S.~M., \& {Rao}, A.~P. 1975, \apss, 37, 275,
  \dodoi{10.1007/BF00640353}

\bibitem[{{Armstrong} \& {Coles}(1978)}]{armstrong1978}
{Armstrong}, J.~W., \& {Coles}, W.~A. 1978, \apj, 220, 346,
  \dodoi{10.1086/155912}

\bibitem[{{Armstrong} {et~al.}(1990){Armstrong}, {Coles}, {Kojima}, \&
  {Rickett}}]{armstrong1990}
{Armstrong}, J.~W., {Coles}, W.~A., {Kojima}, M., \& {Rickett}, B.~J. 1990,
  \apj, 358, 685, \dodoi{10.1086/169022}

\bibitem[{Asai {et~al.}(1998)Asai, Kojima, Tokumaru, Yokobe, Jackson, Hick, \&
  Manoharan}]{asai1998}
Asai, K., Kojima, M., Tokumaru, M., {et~al.} 1998, Journal of Geophysical
  Research: Space Physics, 103, 1991, \dodoi{https://doi.org/10.1029/97JA02750}

\bibitem[{{Baars} {et~al.}(1977){Baars}, {Genzel}, {Pauliny-Toth}, \&
  {Witzel}}]{baars1977}
{Baars}, J.~W.~M., {Genzel}, R., {Pauliny-Toth}, I.~I.~K., \& {Witzel}, A.
  1977, \aap, 61, 99

\bibitem[{{Bagnulo} {et~al.}(2021){Bagnulo}, {Cellino}, {Kolokolova},
  {Ne{\v{z}}i{\v{c}}}, {Santana-Ros}, {Borisov}, {Christou}, {Bendjoya}, \&
  {Devog{\`e}le}}]{bagnulo2021}
{Bagnulo}, S., {Cellino}, A., {Kolokolova}, L., {et~al.} 2021, Nature
  Communications, 12, 1797, \dodoi{10.1038/s41467-021-22000-x}

\bibitem[{{Biermann} {et~al.}(1967){Biermann}, {Brosowski}, \&
  {Schmidt}}]{biermannetal1967}
{Biermann}, L., {Brosowski}, B., \& {Schmidt}, H.~U. 1967, \solphys, 1, 254,
  \dodoi{10.1007/BF00150860}

\bibitem[{{Brandt}(1968)}]{brandt1968}
{Brandt}, J.~C. 1968, \araa, 6, 267,
  \dodoi{10.1146/annurev.aa.06.090168.001411}

\bibitem[{{Celnikier} {et~al.}(1987){Celnikier}, {Muschietti}, \&
  {Goldman}}]{celnikier1987}
{Celnikier}, L.~M., {Muschietti}, L., \& {Goldman}, M.~V. 1987, \aap, 181, 138

\bibitem[{{Cohen} \& {Gundermann}(1969)}]{cohen1969}
{Cohen}, M.~H., \& {Gundermann}, E.~J. 1969, \apj, 155, 645,
  \dodoi{10.1086/149897}

\bibitem[{{Cohen} {et~al.}(1967){Cohen}, {Gundermann}, \& {Harris}}]{cohen1967}
{Cohen}, M.~H., {Gundermann}, E.~J., \& {Harris}, D.~E. 1967, \apj, 150, 767,
  \dodoi{10.1086/149380}

\bibitem[{{Coles}(1978)}]{coles1978}
{Coles}, W.~A. 1978, \ssr, 21, 411, \dodoi{10.1007/BF00173067}

\bibitem[{{Coles}(1995)}]{coles1995}
---. 1995, \ssr, 72, 211, \dodoi{10.1007/BF00768782}

\bibitem[{{Coles} \& {Harmon}(1989)}]{ColesHarmon1989}
{Coles}, W.~A., \& {Harmon}, J.~K. 1989, \apj, 337, 1023,
  \dodoi{10.1086/167173}

\bibitem[{Coles {et~al.}(1978)Coles, Harmon, Lazarus, \&
  Sullivan}]{Colesharmonetal1978}
Coles, W.~A., Harmon, J.~K., Lazarus, A.~J., \& Sullivan, J.~D. 1978, Journal
  of Geophysical Research: Space Physics, 83, 3337,
  \dodoi{https://doi.org/10.1029/JA083iA07p03337}

\bibitem[{{Condon} {et~al.}(1998){Condon}, {Cotton}, {Greisen}, {Yin},
  {Perley}, {Taylor}, \& {Broderick}}]{condon1998}
{Condon}, J.~J., {Cotton}, W.~D., {Greisen}, E.~W., {et~al.} 1998, \aj, 115,
  1693, \dodoi{10.1086/300337}

\bibitem[{{Cordiner} {et~al.}(2020){Cordiner}, {Milam}, {Biver},
  {Bockel{\'e}e-Morvan}, {Roth}, {Bergin}, {Jehin}, {Remijan}, {Charnley},
  {Mumma}, {Boissier}, {Crovisier}, {Paganini}, {Kuan}, \&
  {Lis}}]{cordiner2020}
{Cordiner}, M.~A., {Milam}, S.~N., {Biver}, N., {et~al.} 2020, Nature
  Astronomy, 4, 861, \dodoi{10.1038/s41550-020-1087-2}

\bibitem[{{Douglas} {et~al.}(1996){Douglas}, {Bash}, {Bozyan}, {Torrence}, \&
  {Wolfe}}]{texas1996}
{Douglas}, J.~N., {Bash}, F.~N., {Bozyan}, F.~A., {Torrence}, G.~W., \&
  {Wolfe}, C. 1996, \aj, 111, 1945, \dodoi{10.1086/117932}

\bibitem[{{Gombosi}(2015)}]{gombosi2015}
{Gombosi}, T.~I. 2015, in Geophysical Monograph Series, Vol. 207, Magnetotails
  in the Solar System, ed. A.~{Keiling}, C.~{Jackman}, \& P.~{Delamere},
  169--188, \dodoi{10.1002/9781118842324.ch10}

\bibitem[{{Gordon} {et~al.}(2021){Gordon}, {Boyce}, {O'Dea}, {Rudnick},
  {Andernach}, {Vantyghem}, {Baum}, {Bui}, {Dionyssiou}, {Safi-Harb}, \&
  {Sander}}]{gordon2021}
{Gordon}, Y.~A., {Boyce}, M.~M., {O'Dea}, C.~P., {et~al.} 2021, \apjs, 255, 30,
  \dodoi{10.3847/1538-4365/ac05c0}

\bibitem[{{Hajivassiliou} \& {Duffett-Smith}(1987)}]{1987MNRAS.229..485H}
{Hajivassiliou}, C.~A., \& {Duffett-Smith}, P.~J. 1987, \mnras, 229, 485,
  \dodoi{10.1093/mnras/229.3.485}

\bibitem[{{Harris}(1973)}]{harris1973}
{Harris}, D.~E. 1973, \aj, 78, 369, \dodoi{10.1086/111425}

\bibitem[{{Hewish} {et~al.}(1964){Hewish}, {Scott}, \& {Wills}}]{hewish1964}
{Hewish}, A., {Scott}, P.~F., \& {Wills}, D. 1964, \nat, 203, 1214,
  \dodoi{10.1038/2031214a0}

\bibitem[{{Kojima} \& {Kakinuma}(1987)}]{kojima1987}
{Kojima}, M., \& {Kakinuma}, T. 1987, \jgr, 92, 7269,
  \dodoi{10.1029/JA092iA07p07269}

\bibitem[{{Large} {et~al.}(1981){Large}, {Mills}, {Little}, {Crawford}, \&
  {Sutton}}]{large1981}
{Large}, M.~I., {Mills}, B.~Y., {Little}, A.~G., {Crawford}, D.~F., \&
  {Sutton}, J.~M. 1981, \mnras, 194, 693, \dodoi{10.1093/mnras/194.3.693}

\bibitem[{{Manoharan}(1993)}]{mano1993}
{Manoharan}, P.~K. 1993, \solphys, 148, 153, \dodoi{10.1007/BF00675541}

\bibitem[{{Manoharan}(2009)}]{manolfru2009}
{Manoharan}, P.~K. 2009, in Astronomical Society of the Pacific Conference
  Series, Vol. 407, The Low-Frequency Radio Universe, ed. D.~J. {Saikia}, D.~A.
  {Green}, Y.~{Gupta}, \& T.~{Venturi}, 359

\bibitem[{Manoharan(2010)}]{mano2010a}
Manoharan, P.~K. 2010, in Magnetic Coupling between the Interior and Atmosphere
  of the Sun, ed. S.~Hasan \& R.~J. Rutten (Berlin, Heidelberg: Springer Berlin
  Heidelberg), 324--331

\bibitem[{{Manoharan}(2012)}]{mano2012}
---. 2012, \apj, 751, 128, \dodoi{10.1088/0004-637X/751/2/128}

\bibitem[{{Manoharan} \& {Ananthakrishnan}(1990)}]{mano1990}
{Manoharan}, P.~K., \& {Ananthakrishnan}, S. 1990, Mon. Not. R. Astron. Soc.,
  244, 691

\bibitem[{{Manoharan} {et~al.}(1995){Manoharan}, {Ananthakrishnan}, {Dryer},
  {Detman}, {Leinbach}, {Kojima}, {Watanabe}, \& {Kahn}}]{mano1995SoPh}
{Manoharan}, P.~K., {Ananthakrishnan}, S., {Dryer}, M., {et~al.} 1995, Solar
  Physics, 156, 377, \dodoi{10.1007/BF00670233}

\bibitem[{{Manoharan} {et~al.}(2000){Manoharan}, {Kojima}, {Gopalswamy},
  {Kondo}, \& {Smith}}]{mano2000}
{Manoharan}, P.~K., {Kojima}, M., {Gopalswamy}, N., {Kondo}, T., \& {Smith}, Z.
  2000, Astrophys. J., 530, 1061, \dodoi{10.1086/308378}

\bibitem[{{Manoharan} {et~al.}(1994){Manoharan}, {Kojima}, \&
  {Misawa}}]{manoetal1994JGR}
{Manoharan}, P.~K., {Kojima}, M., \& {Misawa}, H. 1994, \jgr, 99, 23411,
  \dodoi{10.1029/94JA01955}

\bibitem[{{Manoharan} {et~al.}(2021){Manoharan}, {Perillat}, \&
  {AO~Team}}]{manoaas2021}
{Manoharan}, P.~K., {Perillat}, P., \& {AO~Team}. 2021, in American
  Astronomical Society Meeting Abstracts, Vol. 237, American Astronomical
  Society meeting Abstarcts \#237, 405.05

\bibitem[{{Manoharan} {et~al.}(2001){Manoharan}, {Tokumaru}, {Pick},
  {Subramanian}, {Ipavich}, {Schenk}, {Kaiser}, {Lepping}, \&
  {Vourlidas}}]{manoetal2001}
{Manoharan}, P.~K., {Tokumaru}, M., {Pick}, M., {et~al.} 2001, \apj, 559, 1180,
  \dodoi{10.1086/322332}

\bibitem[{{Manzini} {et~al.}(2020){Manzini}, {Oldani}, {Ochner}, \&
  {Bedin}}]{manzinietal2020}
{Manzini}, F., {Oldani}, V., {Ochner}, P., \& {Bedin}, L.~R. 2020, \mnras, 495,
  L92, \dodoi{10.1093/mnrasl/slaa061}

\bibitem[{{Roy} {et~al.}(2007){Roy}, {Manoharan}, \&
  {Chakraborty}}]{royetal2007}
{Roy}, N., {Manoharan}, P.~K., \& {Chakraborty}, P. 2007, \apjl, 668, L67,
  \dodoi{10.1086/522780}

\bibitem[{{Salter} \& {Manoharan}(2019)}]{chris2019}
{Salter}, C., \& {Manoharan}, P.~K. 2019, \baas, 51, 116

\bibitem[{{Slee} {et~al.}(1990){Slee}, {Bobra}, {Waldron}, \& {Lim}}]{slee1990}
{Slee}, O.~B., {Bobra}, A.~D., {Waldron}, D., \& {Lim}, J. 1990, Australian
  Journal of Physics, 43, 801, \dodoi{10.1071/PH900801}

\bibitem[{{Swarup} {et~al.}(1971){Swarup}, {Sarma}, {Joshi}, {Kapahi}, {Bagri},
  {Damle}, {Ananthakrishnan}, {Balasubramanian}, {Bhave}, \&
  {Sinha}}]{swarup1971}
{Swarup}, G., {Sarma}, N.~V.~G., {Joshi}, M.~N., {et~al.} 1971, Nature Physical
  Science, 230, 185, \dodoi{10.1038/physci230185a0}

\bibitem[{Yakovlev \& Pisanko(2018)}]{yakovklev2018}
Yakovlev, O.~I., \& Pisanko, Y.~V. 2018, Advances in Space Research, 61, 552,
  \dodoi{https://doi.org/10.1016/j.asr.2017.10.052}

\bibitem[{{Yamauchi} {et~al.}(1998){Yamauchi}, {Tokumaru}, {Kojima},
  {Manoharan}, \& {Esser}}]{yama1998JGR}
{Yamauchi}, Y., {Tokumaru}, M., {Kojima}, M., {Manoharan}, P.~K., \& {Esser},
  R. 1998, \jgr, 103, 6571, \dodoi{10.1029/97JA03598}

\bibitem[{Zwickl {et~al.}(1988)Zwickl, Hildner, Bame, Gosling, \&
  Sofaly}]{zwickl1988}
Zwickl, R., Hildner, E., Bame, S., Gosling, J., \& Sofaly, K. 1988, EOS, 69,
  1358

\end{thebibliography}

\begin{table}
\begin{center}
  \caption{Occultation Radio Sources Parameters}
  \begin{tabular}{lcccccccccc}
  \hline
  \hline
 Src Name & RA(B1950) & Dec(B1950) & S$_{327}$ & S$_{820}$ & S$_{1400}$ & $\Delta$S$_{327}$/$\Theta$ & Date (2019) 
          & $\epsilon$  & D$_{\rm probe}$ & Obs \\
          & hh:mm:ss.s  & dd:mm:ss.s  & Jy  & Jy  &  Jy  & mJy/mas & (UT range, hh:mm) & (deg)  & km \\
  \hline          
 B0957+142 & 09:57:45.9 & +14:16:00.2 & 3.5 & -- & 1.17 & 700/150 & 20 Oct (11:16-13:55) & 71 & 8$\times10^5$ & AO  \\
 B1019+083 & 10:19:12.5 & +08:23:42.0 & 1.7 & -- & 0.61 & 500/50  & 31 Oct (11:03-13:21) & 64 & 4$\times 10^5$ & AO$^c$ \\   
 B1023+067 & 10:23:55.1 & +06:42:50.5 & 4.3 & -- & 0.85 & 600/100 & 03 Nov (11:02-13:09) & 63 & 8$\times 10^5$ & AO  \\
 B1129--202 & 11:29:44.4 & --20:17:17.8 & 0.8 & 0.37& 0.13 &    --   & 11 Dec (09:15-13:15) & 77 & $\sim10^6$ & GBT$^c$\\
  \hline
\end{tabular}
\end{center}
S$_{327}$ and S$_{1400}$ - source flux densities at 327 and 1400 MHz, 
respectively, obtained from the Ooty Radio Telescope (\citealt{mano2012}) and 
the NRAO VLA Sky Survey (NVSS) (\citealt{condon1998});
S$_{820}$ - source flux density at 820 MHz derived as described in Section 4.3;
$\Delta$S$_{327}$ and $\Theta$ - scintillating flux density and estimated angular
size of the compact component obtained from extensive IPS observations of the Ooty 
Radio Telescope at 327 MHz (\citealt{mano2012}); 
$\epsilon$ - solar elongation of the source; 
D$_{\rm probe}$ - distance between comet nucleus and source line of sight crossing in 
the tail direction;
AO$^c$ and GBT$^c$ - Arecibo and Green Bank observations probed the central part 
of the tail; AO observations made in the 327- and 1400-MHz bands; GBT observations made 
in the 820-MHz band.  
\end{table}

\newpage

\begin{figure}
\centerline{\includegraphics[width=0.50\textwidth,angle=-90]{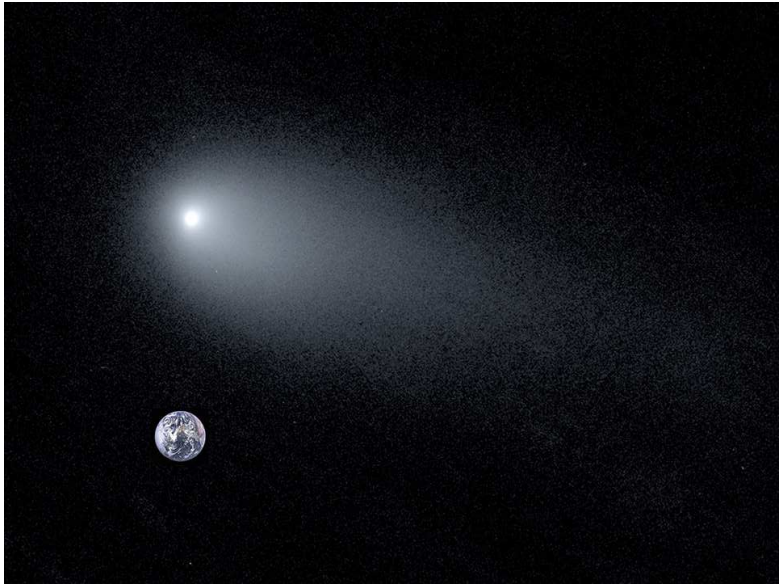}}
\caption{
Image of Comet 2I/Borisov taken by the Keck Observatory's Low-Resolution 
Imaging Spectrometer on 24 November 2019. To illustrate the size scale, 
an image of the Earth is shown (Credit: P. van Dokkum, G. Laughlin, 
C. Hsieh, S. Danieli, Yale University). }
\end{figure}

\newpage 

\begin{figure}
\centerline{\includegraphics[width=0.60\textwidth,angle=-90]{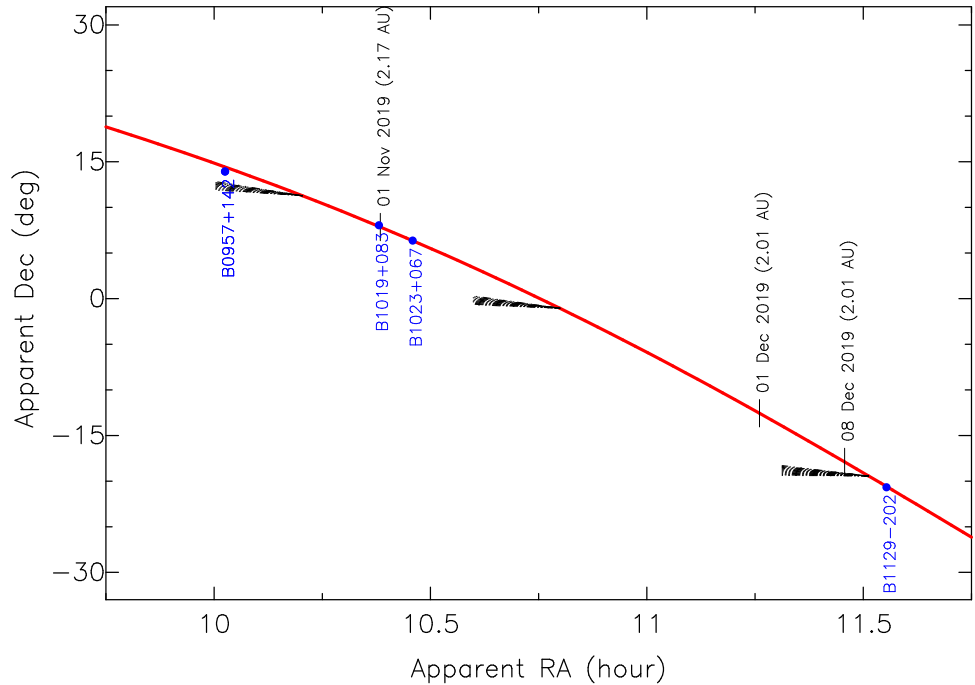}}
\caption{ 
The path of Comet 2I/Borisov from mid-October to mid-December 2019. 
The heliocentric distances of the comet are marked on 01 November, 01 December, 
and on the perihelion date 08 December 2019. Typical tail directions along 
the anti-solar direction are shown. The occultation sources are also plotted along
the path of the comet (see Table 1).  }
\end{figure}

\begin{figure}
\centerline{\includegraphics[width=0.50\textwidth,angle=-90]{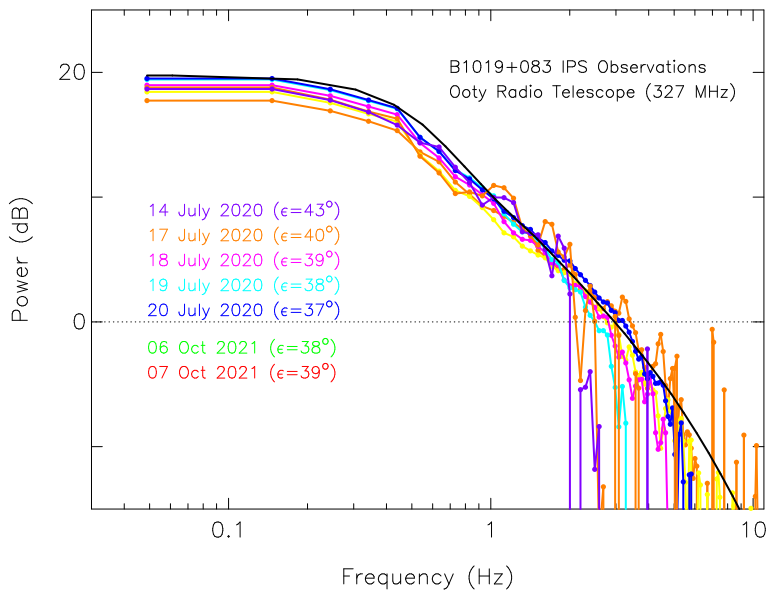}}
\caption{The derived IPS power spectra from observations of B1019+083 after 
subtracting the noise floor. These spectra were obtained from the Ooty Radio Telescope 
at 327 MHz. For each spectrum, the date of observation and the solar elongation 
of the source are shown with the corresponding color code. The horizontal dotted
line indicates the typical subtracted level of the white-noise spectrum due to 
the receiver system noise. A model spectrum for typical solar wind parameters, 
(spectral power-law index, $\alpha$~=~--3.3, and solar wind speed, V = 350 km/s), 
and source size, $\Theta$~=~50~mas, is shown as a continuous curve (solid 
black spectrum).}
\end{figure}

\newpage 

\begin{figure}
\centerline{\includegraphics[width=0.5\textwidth,angle=-90]{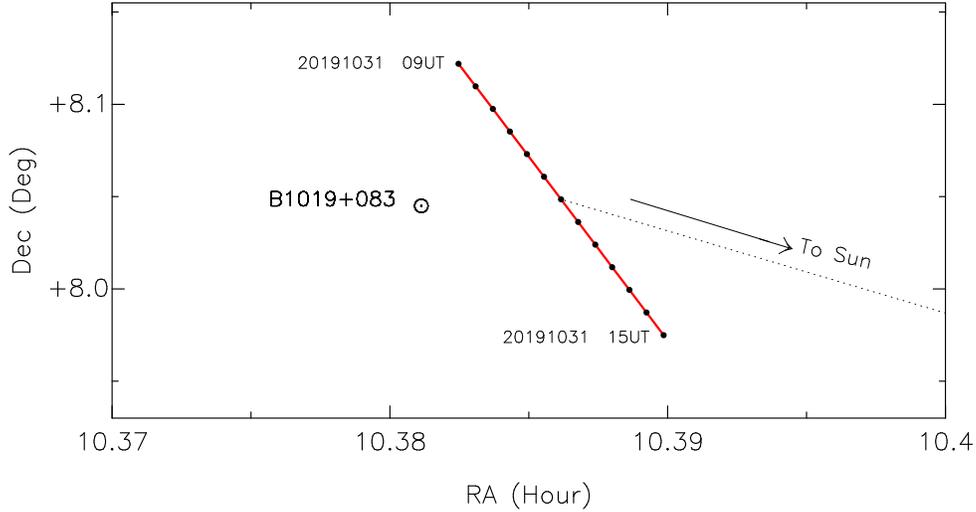}}
\caption{ 
Path of Comet 2I/Borisov on 31 October 2019 (shown in red between 9 and 
15 UT) plotted on the sky plane, with respect to the radio source, B1019+083. 
The plot axes are coordinates on the date of observation. The thin-dotted line 
connects the position of the Sun and the comet nucleus at 12 UT, while the 
arrow indicates the Sunward direction from the comet. As seen in the plot, the 
central portion of the anti-solar comet plasma tail was expected to cross the 
line of sight to B1019+083 around 13 UT.}
\end{figure}

\newpage 

\begin{figure}
\centerline{\includegraphics[width=0.5\textwidth,angle=-90]{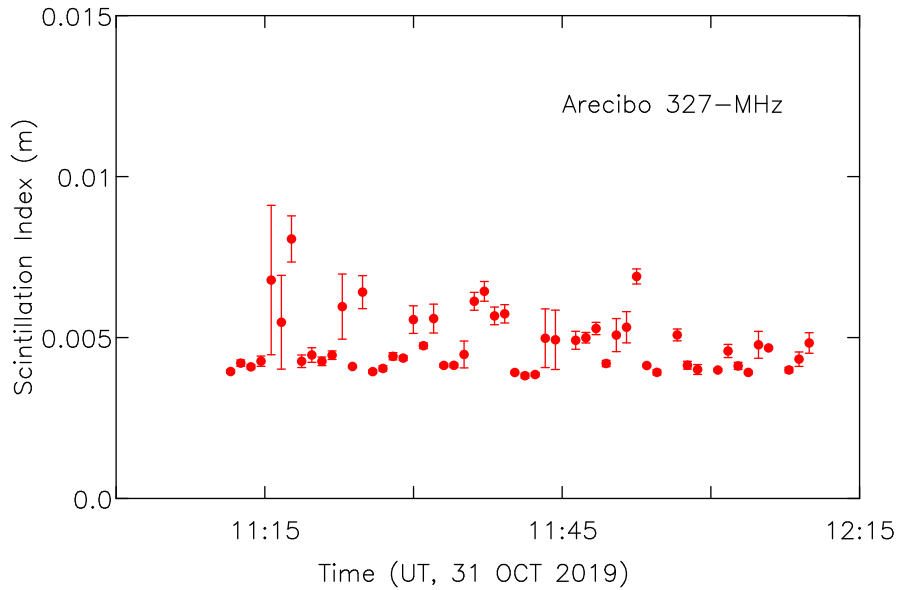}}
\caption{Scintillation indices from the Arecibo observations of 
B1019+083 at 327~MHz on 31 October 2019. This data set was
taken with a bandwidth of $\sim$50 MHz and covers a time interval of about 
11:10 to 12:12 UT. During this time, the line of sight to
the radio source was about 2 arcmin north of the comet tail position. 
Each point on the plot corresponds to the rms computed by integrating
a 1-min temporal power spectrum. 
Each vertical bar represent the `peak-to-peak' variation of 5 scintillation 
indices obtained from the 5 individual subbands of 10-MHz width.
The level of the fluctuations remained essentially constant.}
\end{figure}

\newpage 

\begin{figure}
\centerline{\includegraphics[width=0.52\textwidth,angle=-90]{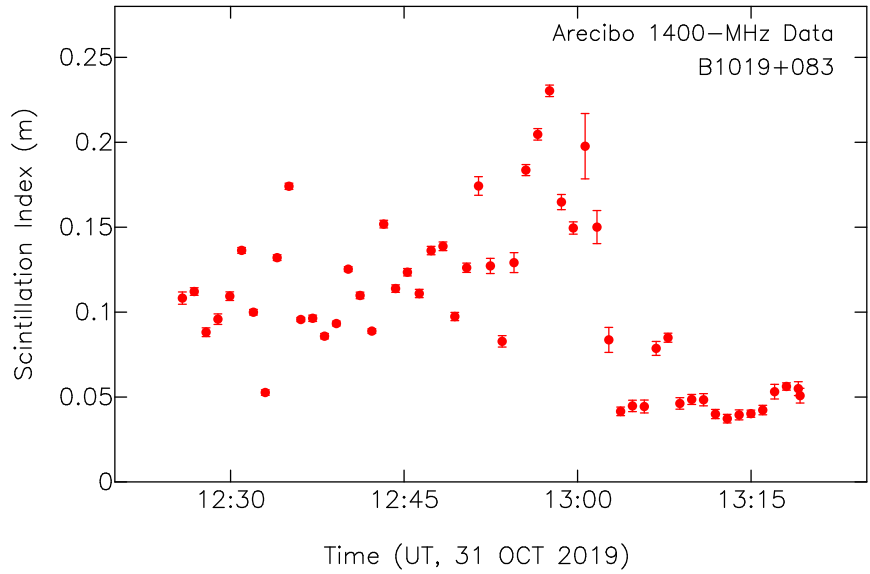}}
\caption{Scintillation indices from Arecibo observations of 
B1019+083 at L-band on 31 October 2019.  Each point 
corresponds to the rms computed by integrating a 1-min temporal power 
spectrum. The vertical bars on the points represent the $\pm$1-$\sigma$ 
uncertainty.
In these observations, a wide bandwidth was employed, which helped generate
highly sensitive temporal power spectra. The central portion of the comet tail
was expected to cross the line of sight around 13 UT.  A gradual increase 
in scintillation was seen from the start of the scan, with the
scintillation peaking around 12:58 UT. From about 13:03 UT, the telescope 
pointing moved off the source and on to ``blank sky'' away
from the target. Thus the plotted points in the time range of 13:04 -- 
13:18 UT represent observations of a region 
that provides a typical ``off-source'' rms level.}
\end{figure}

\newpage 

\begin{figure}
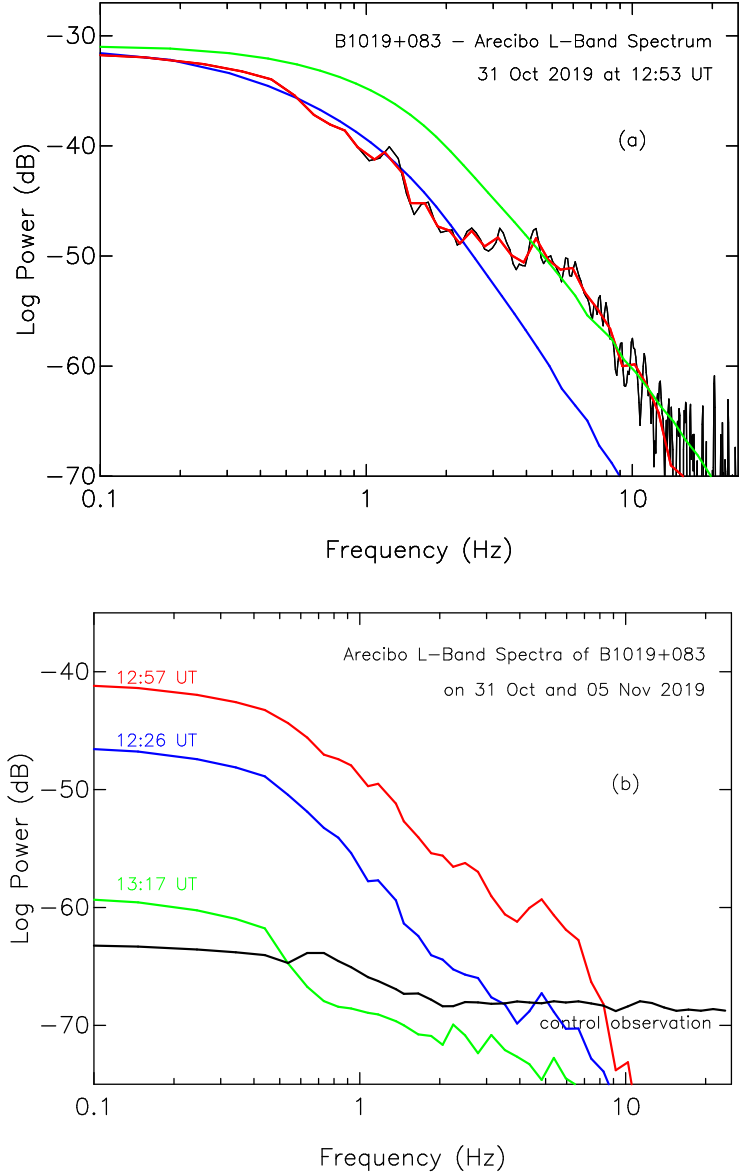

\centerline{\includegraphics[width=7.5cm,angle=-90]{figure_7a.ps}}
\centerline{\includegraphics[width=8.05cm,angle=-90]{figure_7b.ps}}
\caption{
Power spectra of the intensity fluctuations observed on B1019+083 on 
31 October and 05 November 2019.  The horizontal axis is the 
temporal frequency (Hz) and the vertical axis is log power (dB). From each
spectrum, its corresponding white-noise level has been subtracted. 
{\bf (a)}: The observed spectra of B1019+083 at 1400-MHz on 31 October, at 
12:53 UT plotted in `black' and `red' are, respectively, the 
spectrum with an averaging of two adjacent points, and the same spectrum
in which the averaging is such that the points are nearly equispaced on 
the log scale. For comparison, simple model spectra obtained from
Equation~(1) are shown.  The spectrum shown in `green' is for a power-law 
exponent, $\alpha$~=~--2.7 and the spectrum plotted in `blue’ is for 
$\alpha$~=~--3.3. For both model spectra, a spherically symmetric solar-wind 
velocity of 360 km/s, an anisotropy parameter, AR~=~1.5, and a source size 
of 50~mas are used. 
{\bf (b)}: The spectra plotted, in red, blue, and green respectively, 
correspond to 2-min average spectra at $\sim$12:57, $\sim$12:26, and $\sim$13:17
UT (see Figure~6) on 31 October 2019. When the 
scintillation is at the peak, the spectrum is broad and extends up to 
about 10~Hz, and its width/shape provides information on the
speed and size of density irregularities crossing the line of sight to 
the radio source.  The spectrum shown in black corresponds to a control 
observation taken on 05 November 2019, when B1019+083 was well away from 
the tail of the comet at a solar elongation of 69$^\circ$.  It shows the 
typical background rms level. This spectrum is plotted without subtracting 
the white-noise level.
}
\end{figure}

\newpage

\begin{figure}
\centerline{\includegraphics[width=0.5\textwidth,angle=-90]{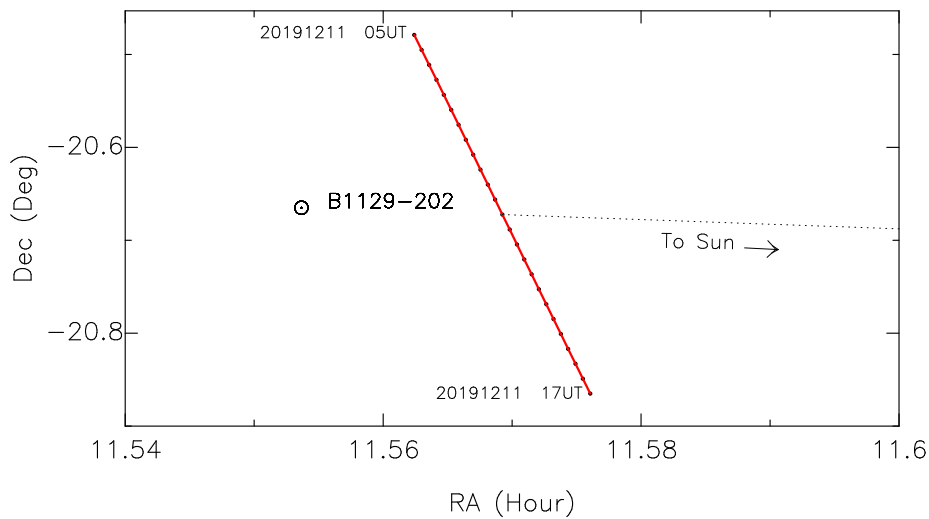}}
\caption{
Similar to Figure 4, the comet's path on 11 December 2019, plotted with 
respect to the radio source B1119-202. The thin-dotted line connects 
the position of the Sun and the comet nucleus at 11 UT. As seen in the plot,
the central portion of the anti-solar comet plasma tail was expected to cross 
the line of sight to B1129-202 around 11 UT.}
\end{figure}

\newpage

\begin{figure}
\centerline{\includegraphics[width=0.5\textwidth,angle=-90]{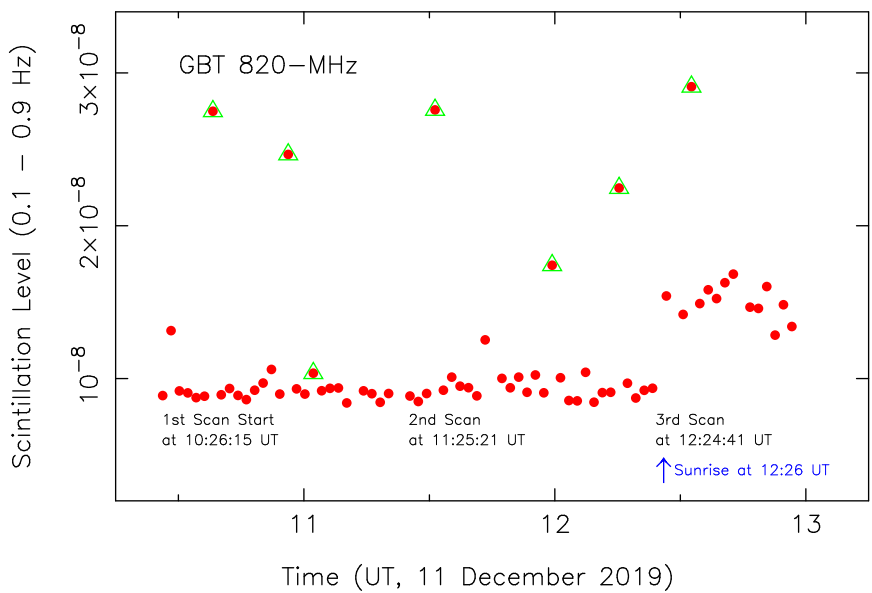}}
\caption{
Relative level of fluctuations from the Green Bank Telescope observations of 
B1129-202 at 820 MHz on 11 December 2019. Each point on the 
plot corresponds to the rms computed by integrating the $\sim$2-min temporal 
power spectrum. The central portion of the comet tail was expected to cross 
the line of sight around 11:00 UT. However, we observed a nearly steady level of 
fluctuation up to the sunrise at $\sim$12:26 UT, following which an increase in 
the level of scintillation was observed. The points enclosed by the `green'
triangles are likely affected by interference. }
\end{figure}

\begin{figure}
\centerline{\includegraphics[width=7.5cm,angle=-90]{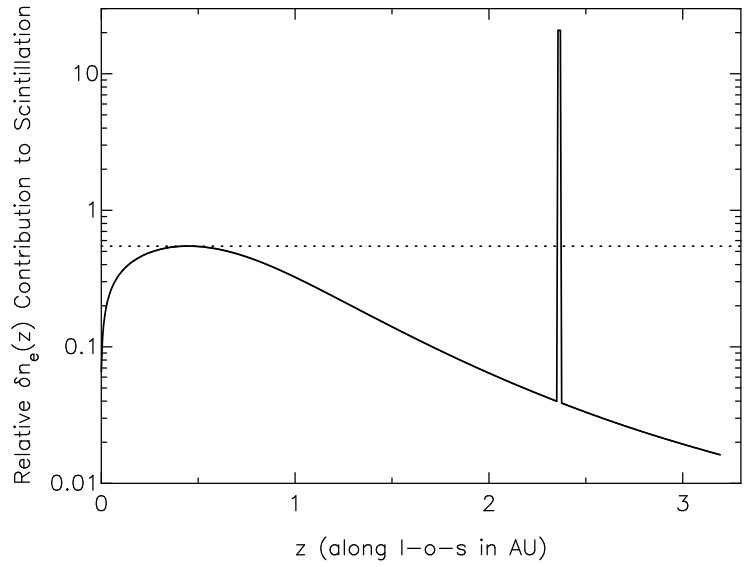}}
\caption{
The rms electron density contribution of the background solar wind to the level 
of scintillation at 1420 MHz along the line-of-sight between the observer and a 
distance of 3.2 AU. The line-of-sight crosses the plasma tail of the comet at 
2.4~AU. See Section~5 for details.
}
\end{figure}

\end{document}